\documentclass[usenatbib]{mn2e}

\usepackage{graphicx}
\usepackage{mathptmx}
\usepackage[hang,raggedright]{subfigure}
\usepackage{amssymb}
\usepackage{amsmath}
\graphicspath{{./Images/}}
\begin{document}

\title[Dust extinction and stellar mass]{Predicting dust
  extinction from the stellar mass of a galaxy}

\author[T.\ Garn et al.]{Timothy Garn\thanks{Deceased}, Philip
  N.\ Best\thanks{Email: pnb@roe.ac.uk}
\\   
    SUPA, Institute for Astronomy, Royal Observatory Edinburgh,
    Blackford Hill, Edinburgh, EH9~3HJ, UK\\
}

\date{\today}
\pagerange{\pageref{firstpage}--\pageref{lastpage}; } \pubyear{2010}
\label{firstpage}
\maketitle

\begin{abstract}
We investigate how the typical dust extinction of H$\alpha$ luminosity
from a star-forming galaxy depends upon star formation rate (SFR),
metallicity and stellar mass independently, using a sample of $\sim$90,000
galaxies from Data Release 7 of the Sloan Digital Sky Survey (SDSS).  We
measure extinctions directly from the Balmer decrement of each source, and
while higher values of extinction are associated with an increase in any
of the three parameters, we demonstrate that the fundamental property that
governs extinction is stellar mass.  After this mass-dependent
relationship is removed, there is very little systematic dependence of the
residual extinctions with either SFR or metallicity, and no significant
improvement is obtained from a more general parameterisation.  In contrast
to this, if either a SFR-dependent or metallicity-dependent extinction
relationship is applied, the residual extinctions show significant trends
that correlate with the other parameters.  Using the SDSS data, we present
a relationship to predict the median dust extinction of a sample of
galaxies from its stellar mass, which has a scatter of $\sim0.3$~mag.  The
relationship was calibrated for H$\alpha$ emission, but can be more
generally applied to radiation emitted at other wavelengths.  These
results have important applications for studies of high-redshift galaxies,
where individual extinction measurements are hard to obtain but stellar
mass estimates can be relatively easily estimated from long-wavelength
data.

\end{abstract}


\section{Introduction}
\label{sec:introduction}

Galaxies which are currently forming stars emit a large fraction of
their luminosity at ultraviolet (UV) wavelengths, where the output
from young, massive stars peaks.  Some fraction of this radiation is
absorbed and re-emitted through the H$\alpha$ recombination line, and
the intrinsic luminosity of both of these indicators scales directly
with the rate of formation of massive stars ($\gtrsim 10$~$M_{\odot}$,
with lifetimes less than $\sim20$~Myr), giving a sensitive,
well-calibrated probe of the current star formation rate \citep[SFR;
e.g.][]{Kennicutt98}.  However, interstellar dust grains attenuate
this short-wavelength radiation, reducing the observed flux and
re-processing it into the far-infrared, so that only a small fraction
of the intrinsic H$\alpha$ or UV luminosity of a galaxy can be
measured.  Studies of star-forming galaxies are therefore hindered by
uncertainties about the nature and amount of dust extinction which has
occurred.

The typical H$\alpha$ extinction of a galaxy, $A_{\rm H\alpha}$, can
vary by several magnitudes and has been shown to depend upon galaxy
properties such as luminosity \citep[e.g.][]{Wang96}, SFR
\citep[e.g.][]{Hopkins01, Sullivan01, Berta03, Garn09extinction},
stellar mass \citep[e.g.][]{Brinchmann04, Stasinska04, Pannella09} and
metallicity \citep[e.g.][]{Heckman98, Boissier04, Asari07}.
Explanations have been proposed for all of these effects: more massive
galaxies may have built up a larger dust reservoir with which to
attenuate H$\alpha$ radiation; more actively star-forming galaxies are
likely to have larger and more dusty star-forming regions; more
metal-rich galaxies have a greater dust-to-gas ratio, and therefore
higher extinction.

However, these studies have typically focused on one dependence at a time,
and while they confirm that a single value of extinction should not be
assumed for all galaxies, they do not address the question of whether
stellar mass, SFR or metallicity is more fundamental.  Further
complications are introduced through the known correlations between galaxy
properties -- more massive galaxies tend to be more metal-rich
\citep[e.g.][]{Tremonti04, Kewley08} and, amongst star-forming galaxies,
more massive galaxies also have a greater typical SFR
\citep[e.g.][]{Brinchmann04}\footnote{The mass-SFR correlation breaks down
  for masses above those at which `SFR-downsizing' has set in
  \citep[e.g.][]{Cowie96}, whereby the most massive galaxies are no longer
  star-forming in the nearby Universe.}. Recently \citet{Mannucci10} have
proposed the existence of a `fundamental metallicity relation', whereby
galaxies define a tight surface in the 3-dimensional space of mass, SFR
and metallicity. They also argue that this surface does not evolve out to
$z \sim 2.5$, and that previous observations that the trends between mass,
metallicity and SFR evolve with redshift
\citep[e.g.][]{Erb06,Daddi07,Maiolino08} arise because the high-redshift
observations sample different regions of the surface.  What is certainly
clear is that the large number of competing effects make it difficult to
determine what drives the observed systematic variation in dust extinction
between different galaxies.

The most direct method of estimating extinction makes use of the strong
wavelength-dependence of dust attenuation -- by comparing the observed
flux ratio of the H$\alpha$ and H$\beta$ lines to the ratio that is
expected in the absence of dust (the `Balmer decrement'), the amount of
extinction at a given wavelength can be calculated.  However, this method
is only practical at low redshift, given the relative weakness of the
H$\beta$ line, and the fact that the H$\alpha$ line is redshifted out of
the optical band at $z\gtrsim0.5$.  If a suitable parameterisation for
extinction could be calculated from galaxies in the local Universe, it
could usefully be applied to the high-redshift population, where
extinction cannot be estimated as easily.

In this work we aim to disentangle the effects that stellar mass, SFR
and metallicity have upon dust extinction, determine which of these
parameters is the most fundamental, and define a relationship to
estimate the typical extinction of a galaxy when measurements of the
Balmer decrement are not available.  In Section~\ref{sec:sample} we
present the data, describe our methods for selecting a clean sample of
star-forming galaxies, and describe the estimates of SFR, stellar
mass, metallicity and extinction that we use.
Section~\ref{sec:results} presents the observed variation in H$\alpha$
extinction with each combination of galaxy parameters, and our models
that describe how to predict dust extinction.  In
Section~\ref{sec:discussion} we discuss these results, and their
implications.

Throughout this work we assume a concordance cosmology of $\Omega_{\rm
M}=0.3$, $\Omega_{\Lambda}=0.7$ and H$_{0}=70$~km~s$^{-1}$~Mpc$^{-1}$.

\section{Data sample}
\label{sec:sample}
\subsection{Selection of a star-forming sample}
In this work we make use of the superb spectroscopic information
available from the Sloan Digital Sky Survey, Data Release~7
\citep[SDSS DR7;][]{Abazajian09}.  This dataset has been analysed by a
group from the Max Planck Institute for Astrophysics, and Johns
Hopkins University (hereafter the MPA/JHU group)\footnote{
http://www.mpa-garching.mpg.de/SDSS/DR7/} -- descriptions of the
analysis pipeline for a previous SDSS data release can be found in
\citet{Brinchmann04} and \citet{Tremonti04}.  Spectroscopic redshifts
and line fluxes (from a Gaussian fit to continuum-subtracted data,
corrected for Galactic reddening) are provided for 927,552 sources at
$z<0.7$.  We exclude from our analysis the sources spectroscopically
classified as a `QSO', sources outside the redshift range
$0.04<z<0.2$, within the redshift range $0.146 < z < 0.148$ (see
Section~\ref{sec:extinction}), or sources with an uncertain estimate
of their redshift (Z\_WARNING $>0$), and also remove multiple entries
of a number of objects from the dataset (where we define a multiple
entry to consist of two catalogued sources with angular separation
less than 5~arcsec, and $\Delta z < 0.001$).  617,822 sources are
retained after these criteria have been applied.

In this work we require a sample of star-forming galaxies, from which we
can measure an extinction and metallicity from emission-line ratios.  In
order to distinguish between star-forming galaxies and AGN, we use the
`BPT' \citep*{Baldwin81} emission line diagnostic, which compares the
[{\sc O\,iii}]5007/H$\beta$ and [{\sc N\,ii}]6584/H$\alpha$ line flux
ratios.  However, in order to use this diagnostic, all four emission lines
must be detected with sufficient significance that accurate line flux
ratios can be calculated.  Note that the formal line flux uncertainties
quoted in the MPA/JHU DR7 catalogue significantly underestimate the true
errors -- we have increased these uncertainties by the recommended factors
of 2.473, 1.882, 1.566 and 2.039 for the H$\alpha$, H$\beta$, [{\sc
    O\,iii}] and [{\sc N\,ii}] lines respectively, which were calculated
by the MPA/JHU group through comparisons of the derived line fluxes of
sources that were observed multiple times.

\begin{figure}
  \begin{center}
    \includegraphics[width=0.4\textwidth]{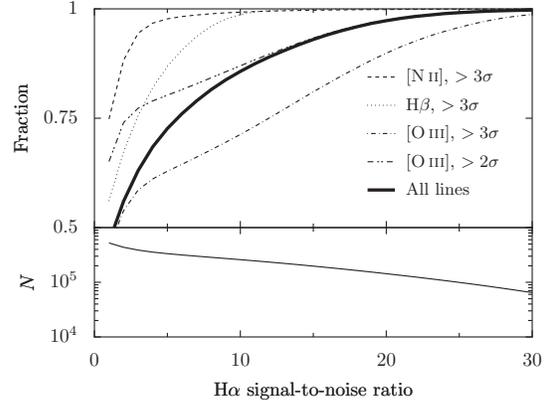}
    \caption{Lower panel: the number of sources above a given H$\alpha$
      S/N ratio.  Upper panel: the fraction of these sources that have a
      detection of the H$\beta$, [{\sc O\,iii}] or [{\sc N\,ii}] lines
      above a given S/N ratio, and the fraction with detections of all
      three lines, with S/N of $>3$ for the H$\beta$ and [{\sc N\,ii}]
      lines, and S/N of $>2$ for the [{\sc O\,iii}] line.}
    \label{fig:emissionlines}
  \end{center}
\end{figure}

We perform a primary selection on the signal-to-noise (S/N) of the
H$\alpha$ line (we investigate possible biases of this selection method in
Section~3.5), and measure the fraction of these sources that also have a
detection of the other lines with a S/N of $>3$.
Fig.~\ref{fig:emissionlines} shows how the number of sources above a given
H$\alpha$ S/N ratio varies, along with the fraction of these sources with
detections of the other emission lines above a given significance.  It is
clear that the limiting factor in this analysis is obtaining a secure
detection of the [{\sc O\,iii}] line -- any sample with an H$\alpha$ S/N
threshold $\gtrsim 10$ is essentially an [{\sc O\,iii}]-selected sample.

For this analysis we set an H$\alpha$ S/N requirement of 20, which
includes 143,472 sources.  We set an H$\beta$ S/N requirement of 3, which
includes 99.97~per~cent of these (i.e.\ we are not biasing the sample
through our H$\beta$ cut).  Of these sources, 99.86~per~cent had a
detection of the [{\sc N\,ii}] line with S/N $>3$.  In order to not bias
our sample against weak [{\sc O\,iii}] emitters (which have low
metallicity; see Section~\ref{sec:properties}), we reduce the [{\sc
    O\,iii}] S/N threshold to 2.  This cut includes 139,586 sources
(97~per~cent), in comparison to the 89~per cent which would have been
selected above a S/N of 3.  The \citet{Kauffmann03BPT} criteria was then
used to identify sources on the BPT diagram which are definitely
star-forming (120,650 galaxies; 86~per~cent of the sample).

\subsection{Sample properties}
\label{sec:properties}
\subsubsection{Extinction}
\label{sec:extinction}
For each galaxy we measure the H$\alpha$ to H$\beta$ line flux ratio,
$S_{\rm H\alpha}/S_{\rm H\beta}$, and estimate an extinction at a
particular wavelength $\lambda$, $A_{\lambda}$, using
\begin{equation}
A_{\lambda} = \frac{-2.5 k_{\lambda}}{k_{\rm H\beta}-k_{\rm
    H\alpha}}~{\rm log}_{10}\left(\frac{2.86}{S_{\rm H\alpha}/S_{\rm
    H\beta}}\right),
\label{eq:extinction}
\end{equation}
where 2.86 is the intrinsic H$\alpha$/H$\beta$ line flux ratio that is
appropriate for Case~B recombination, a temperature of $T = 10^{4}$~K
and an electron density of $n_{\rm e} = 10^{2}$~cm$^{-3}$
\citep{Brocklehurst71}.  We consider the accuracy of using the single
value of 2.86 for all galaxies in Section~\ref{sec:accuracy}.

We use the \citet{Calzetti00} dust attenuation law to calculate the
values of $k_{\lambda} \equiv A_{\lambda} / E(B-V)$ at the wavelengths
of the H$\alpha$ and H$\beta$ emission lines.  This relationship was
measured from observations of $z\sim0$ starburst galaxies, and has
been shown to be appropriate for galaxies at redshifts up to $\sim0.8$
\citep{Garn09extinction}.  Galaxies which had an observed value of
$S_{\rm H\alpha}/S_{\rm H\beta}<2.86$ were assigned an extinction of
0~mag (0.6~per~cent of the sample).

An uncertainty in the extinction for each galaxy was calculated, using the
H$\alpha$ and H$\beta$ flux uncertainties and standard error propagation
methods; the median uncertainty is 0.2 mag. The extinction errors were not
found to correlate with any of the other properties of the sample,
although a significant increase in the typical error was found for
galaxies within the redshift range $0.146 < z < 0.148$ (median uncertainty
of 0.34).  This appears to be due to contamination of the redshifted
H$\beta$ line with the [{\sc O\,i}]5577 sky emission line
\citep{Osterbrock92} -- we reject from our analysis any sources within
this redshift range.

\subsubsection{Stellar mass}
Stellar mass estimates, $M_{*}$, have been calculated for the DR7 data by
the MPA/JHU
team\footnote{http://www.mpa-garching.mpg.de/SDSS/DR7/Data/stellarmass.html},
based upon fits to the total (CMODEL) SDSS photometry and following the
philosophy of \citet{Kauffmann03stellarmass} and \citet{Salim07}. We adopt
these mass estimates, and we exclude from our analysis all galaxies
(7~per~cent of the remaining sample) where this estimate was not
available. Note that the excluded galaxies have comparable properties to
those retained, so this exclusion should not bias the results in any way.

Additional stellar mass estimates have been calculated from photometry
measured within the 3~arcsec diameter SDSS fibres.  A comparison of
these two estimates allows us to define an effective aperture
correction factor, which corresponds to the fraction of the stellar
mass (and hence broadly, the stellar light) that is located within the
fibre -- this correction factor can be used to identify and reject
highly extended galaxies, where the fibre-based spectroscopy will not
be representative of the properties of the whole galaxy.  This is
effectively the same correction as is discussed by
\citet{Hopkins03SDSS}, who use the ratio of total and fibre-based
$r$-band magnitudes to correct the detected H$\alpha$ emission of SDSS
galaxies.

\citet{Kewley05} discuss the potential effects of estimating Balmer
extinctions, metallicities and SFRs from spectroscopic observations made
within a limited aperture, and conclude that an aperture covering factor
of 20~per~cent is sufficient to minimize systematic differences between
these quantities and the global values.  They recommend a minimum redshift
of $z=0.04$ in order to obtain an average fibre-covering factor of greater
than 20~per~cent for a typical galaxy, which we have adopted for this
analysis.  We further reject galaxies at higher redshift that have a
fibre-coverage factor $<20$~per~cent, along with a small number of
galaxies where a conversion between point spread function magnitudes and
fibre magnitudes performed by the MPA/JHU group has been unsuccessful
(J.~Brinchmann, private communication), leaving 95,240 galaxies in the
sample (85~per~cent of the galaxies which had stellar mass information
available).

\subsubsection{Star formation rate}
In order to estimate SFRs, we calculate the intrinsic H$\alpha$
luminosity for each galaxy, using the aperture and
extinction-corrected H$\alpha$ line flux.  From this luminosity we
calculate a SFR using the relationship in \citet{Kennicutt98},
modified slightly to assume a \citet{Kroupa01} Initial Mass Function
(IMF) that extends between 0.1 and 100~$M_{\odot}$ (to agree with the
IMF used in the MPA/JHU stellar mass calculations).  The choice of IMF
is unimportant in this analysis as H$\alpha$ emission is produced
mainly by high-mass stars, where most IMFs agree.  Implicit in our
calculation of SFRs from an aperture-corrected H$\alpha$ flux is the
assumption that the radial distributions of H$\alpha$ flux and stellar
continuum are comparable -- both \citet{Brinchmann04} and
\citet{Kewley05} find that this is an acceptable approximation, if at
least 20~per~cent of the light of a galaxy is contained within the
fibre.

\subsubsection{Metallicity}
We use the O3N2 indicator \citep{Pettini04} as our proxy for
`metallicity' (denoting the gas-phase abundance of oxygen relative to
hydrogen), where
\begin{equation}
{\rm O3N2} \equiv {\rm
    log}_{10}\left(\frac{{\rm [O_{III}]5007/H\beta}}{{\rm
    [N_{II}]6584/H\alpha}}\right).
\end{equation}
We use this indicator for three reasons: (i) due to the similarity in
wavelengths for the H$\alpha$ and [{\sc N\,ii}] lines, and the
H$\beta$ and [{\sc O\,iii}] lines, this indicator is essentially
independent of the effects of dust attenuation; (ii) the O3N2
indicator has a unique metallicity for each line flux ratio
\citep[see][for further details]{Pettini04}; (iii) the emission lines
that are required to construct the O3N2 diagnostic are the same as we
have used to select star-forming galaxies from the BPT diagram, and we
can therefore measure a metallicity for our entire sample, without
introducing additional selection criteria.

The conversion to an $({\rm O/H})$ abundance comes from
\citet{Pettini04}, $12 + {\rm log}_{10}({\rm O/H}) = 8.73 - 0.32\times
{\rm O3N2}$, where they caution that the relationship may break down
for low-metallicity galaxies with O3N2 $>1.9$, or $12 + {\rm
log}_{10}({\rm O/H}) < 8.1$ -- only 0.2~per~cent of our sample have
these values of O3N2, and we reject these from our analysis.  All
$({\rm O/H})$ abundances can be converted to solar metallicities,
using a value of $12 + {\rm log}_{10}({\rm O/H})_{\odot} = 8.66$
\citep{Asplund04}, and the range $8.1 < 12 + {\rm log}_{10}({\rm O/H})
< 9$ corresponds to $0.28 < Z/Z_{\odot} < 2.19$.

\subsubsection{Summary}

\begin{table}
\caption{\label{tabsel}Details of the selection cuts applied to the SDSS
  sample.}
\begin{tabular}{lrr}
Reason &  N$_{gals}$ retained & \% removed by cut \\
\hline
Initial Sample                   &  927,552 &  -- \\
Restrict to $0.04<z<0.20$        &  666,978 & 28.09\% \\
Reject $0.146<z<0.148$           &  659,737 &  1.09\% \\
Reject uncertain $z$             &  658,464 &  0.19\% \\
Reject QSOs                      &  657,321 &  0.17\% \\
Remove multiple entries          &  617,822 &  6.01\% \\
Require S/N (H$\alpha$)$>$20     &  143,472 & 76.78\% \\
Require S/N (H$\beta$)$>$3       &  143,433 &  0.03\% \\
Require S/N ([NII])$>$3          &  143,237 &  0.14\% \\
Require S/N ([OIII])$>$2         &  139,586 &  2.55\% \\
Select as SF gals using BPT      &  120,650 & 13.57\% \\ 
Reject if no mass estimate       &  112,008 &  7.16\% \\
Reject if $<$20\% fibre coverage &   95,240 & 14.97\% \\
Restrict to O3N2 $>$ 1.9         &   95,041 &  0.21\% \\ 
\end{tabular}
\end{table}
 
The details of the selection cuts applied to the sample are summarised in
Table~\ref{tabsel}. The vast majority of the selection cuts are either
bias-free (e.g. cuts in redshift), or remove so few galaxies that any bias
must be insignificant. The possible exceptions to this are the selection
of star-forming galaxies using a signal-to-noise threshold in H$\alpha$,
the removal of galaxies without stellar mass estimates, and the
fibre-coverage cut. As noted in Section 2.2.2, the properties
(ie. magnitudes, redshifts, $H\alpha$ fluxes, emission line ratios, etc)
of the galaxies excluded due to an absence of a mass estimate show a
comparable distribution to those of the retained galaxies, so this cut is
not expected to introduce any significant bias. The fibre coverage cut
removes galaxies with larger angular sizes; these tend to be at the lower
end of the redshift range, and to be amongst the more massive galaxies at
those redshifts. This population of galaxies is well-represented at higher
redshifts in the sample (where their angular sizes are smaller, but they
remain luminous enough to be detected), so again their exclusion is
expected to introduce only minimal bias. We conclude that the only
selection cut which has the potential to introduce a significant bias is
that of the H$\alpha$ signal-to-noise threshold, and we will investigate
this further in Section~3.5.

\begin{figure}
  \centerline{\includegraphics[width=0.4\textwidth]{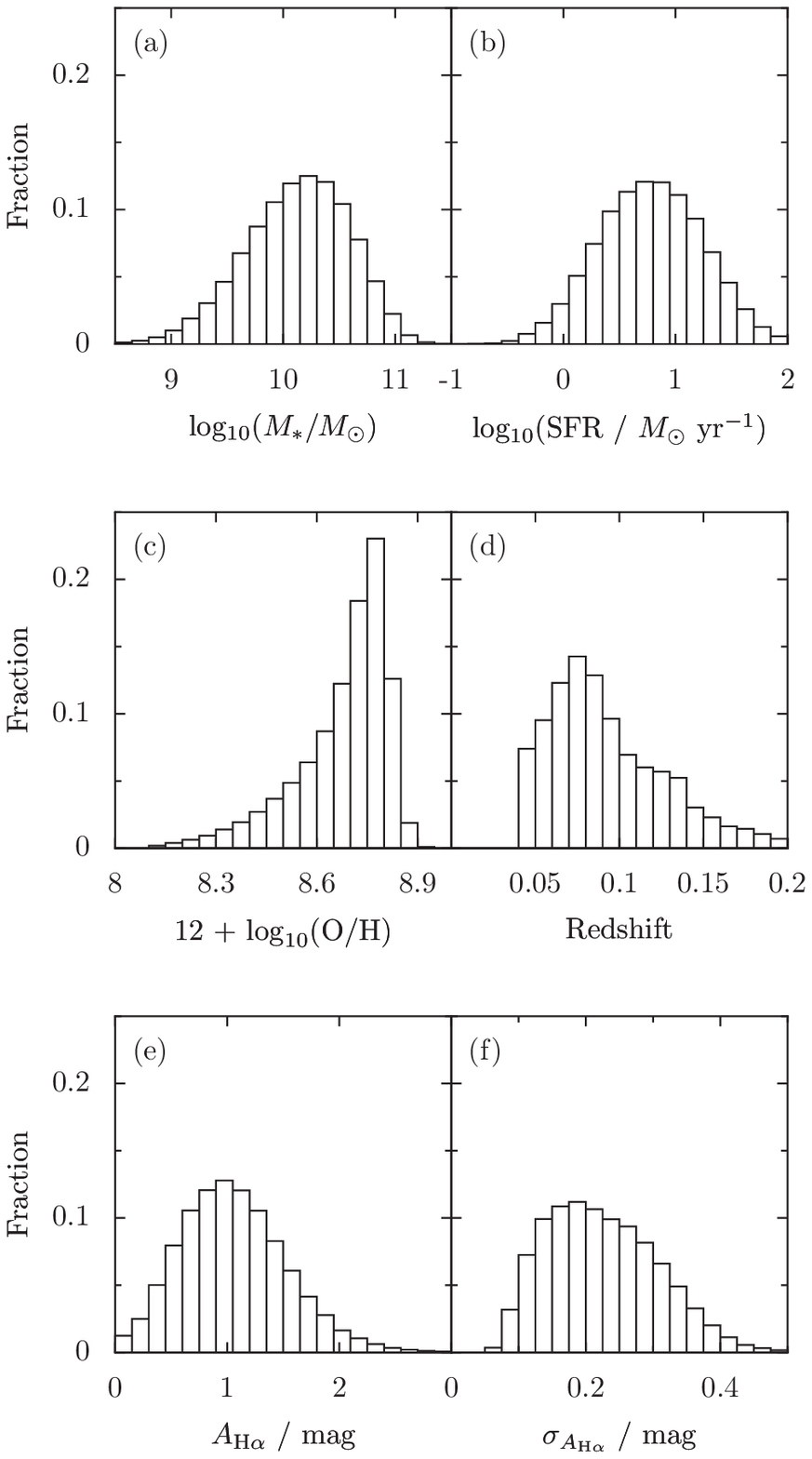}}
  \caption{The fractional distribution of (a) stellar mass; (b)
    star-formation rate; (c) metallicity; (d) redshift; (e) H$\alpha$
    extinction and (f) uncertainty in H$\alpha$ extinction for our
    sample.}
  \label{fig:properties}
\end{figure}

The final number of galaxies in our sample is 95,041, and
Fig.~\ref{fig:properties} shows the distributions of $M_{*}$, SFR,
metallicity, redshift, extinction and extinction uncertainty.  The median
stellar mass is $10^{10.16}$~$M_{\odot}$, with an inter-quartile range of
$10^{9.82}$~$M_{\odot}$ to $10^{10.47}$~$M_{\odot}$. The median SFR is
6.1~$M_{\odot}$~yr$^{-1}$, with an inter-quartile range of 2.9 to
12.9~$M_{\odot}$~yr$^{-1}$. The median metallicity is $12 + {\rm
  log}_{10}({\rm O/H}) = 8.72$, or $Z = 1.1~Z_{\odot}$; the metallicity
distribution is asymmetric with an inter-quartile range of 8.61 to 8.78,
and a longer tail towards low metallicities.  The sample was restricted to
a redshift range of $0.04<z<0.2$ and is mainly at $z<0.1$ (median = 0.085;
inter-quartile range 0.067 to 0.113). The galaxies in the higher redshift
tail predominantly have high masses and high SFR, but their extinction
properties are comparable to those of lower redshift galaxies of the same
mass and SFR -- there is no evidence for redshift evolution within the
sample. The median H$\alpha$ dust extinction is 1.03~mag, close to the
canonical value of 1~mag often assumed for H$\alpha$-selected galaxies.
The median uncertainty in extinction estimates is 0.2~mag.
 
\subsection{Accuracy of the extinction estimates}
\label{sec:accuracy}
\begin{figure*}
  \begin{center}
    \includegraphics[width=0.8\textwidth]{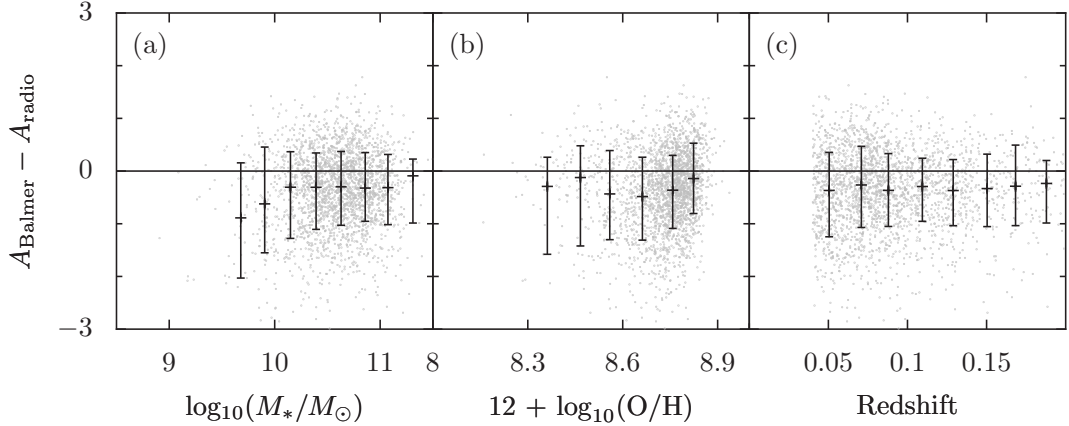}
  \end{center}
  \caption{The difference in extinction estimates between those calculated
    from the Balmer decrement, $A_{\rm Balmer}$, and those from a
    comparison with radio-based SFRs, $A_{\rm radio}$, as a function of
    (a) stellar mass; (b) metallicity and (c) redshift.  Individual
    galaxies are denoted by dots, and binned data (median, $\pm$~$1\sigma$)
    by points with error bars.}
  \label{fig:radioextinction}
\end{figure*}
A previous SDSS data release has been analysed by the MPA/JHU group in
the same manner as the DR7 data which we use in this paper.
\citet{Brinchmann04} have used these data to test the two assumptions
that we have made, namely that the intrinsic H$\alpha$/H$\beta$ ratio
is equal to 2.86, and that the conversion between intrinsic H$\alpha$
luminosity and SFR that we use is appropriate for all galaxies.  They
conclude that both assumptions are suitable when used in combination
with each other, although each is slightly incorrect -- the Case~B
ratio depends upon temperature and density, varying between extremes
of 2.72 (for $T = 2\times10^{4}$~K, $n_{\rm e} = 10^{6}$~cm$^{-3}$)
and 3.03 ($T = 5\times10^{3}$~K, $n_{\rm e} = 10^{2}$~cm$^{-3}$),
while the most metal-rich galaxies produce a slightly lower H$\alpha$
luminosity for the same SFR.  \citet{Brinchmann04} conclude that the
combination of the two assumptions approximately cancel out when
calculating the SFR, but there is the potential for a bias in our
extinction estimates if the wrong intrinsic line ratio has been
assumed.  However, even the maximum error in our choice of Case~B
ratio would only lead to an error in extinction of $\pm0.16$~mag,
which is smaller than the typical uncertainty in the individual
estimates of extinction.

In order to look for systematic offsets in our estimate of extinction
with other properties of a galaxy, we match our SDSS sample with the
1.4-GHz Faint Images of the Radio Sky at Twenty-cm
\citep[FIRST;][]{Becker95} survey, using the methodology described in
\citet{Best05SDSS}.  2404 galaxies in our sample have a unique
counterpart in FIRST, and for each galaxy we calculate a radio-based
SFR using the relationship in \citet{Bell03}, assuming that the radio
spectrum of each galaxy can be modelled by $S_{\nu} \propto
\nu^{-0.8}$ \citep[see e.g.][]{Garn08LH}.  This SFR indicator has been
shown to agree with infrared-based values for star-forming galaxies at
redshifts up to $z=2$ \citep{Garn09SFR}, and gives an independent
estimate of the intrinsic SFR, without being biased by the effects of
dust extinction.

The resolution of FIRST is 5~arcsec, so the radio flux will be
sampling a slightly larger, but comparable region of each galaxy to
the SDSS fibres.  We therefore compare the radio-based SFRs to SFRs
calculated from the observed H$\alpha$ flux, without correcting for
extinction or aperture size, in order to obtain an estimate of the
H$\alpha$ dust extinction for each galaxy, $A_{\rm radio}$, that is
independent of the value calculated from the Balmer decrement, $A_{\rm
Balmer}$.  Fig.~\ref{fig:radioextinction}(a) shows the agreement
between these two extinction estimates -- data have been binned by
stellar mass, and for each bin we calculate the distribution of the
value of $\Delta A \equiv A_{\rm Balmer} - A_{\rm radio}$.  The median
value of $\Delta A$ is taken to be our indicator of the `typical'
value, and the values of $\Delta A$ at the 16th and 84th percentiles
($\pm1\sigma$) to indicate the range that each bin covers.

A systematic offset is seen for all bins, which is equivalent to having
the radio SFRs systematically 30~per~cent greater than the
extinction-corrected H$\alpha$ SFRs.  This offset is likely to be due to
the fact that the FIRST and SDSS observations are estimating their SFRs
from different portions of a galaxy, although there may also be some
contribution from a systematic uncertainty in the two SFR calibrations.
However, no significant mass-dependent variation is seen between the two
estimates of extinction.  Fig.~\ref{fig:radioextinction}(b) and (c) show
the equivalent plots for galaxies binned by metallicity and redshift --
again, no significant variation is seen in the two extinction estimates.
We do not show the equivalent plot for data binned by SFR, as this is
affected significantly by sample bias due to the FIRST completeness limit
of 0.75~mJy corresponding to a SFR limit that is significantly higher than
that of the H$\alpha$ selection.

The results of this section lead us to conclude that our extinction
estimates are robust, and that our choice of a single value of the
intrinsic H$\alpha$/H$\beta$ ratio should not introduce mass,
metallicity or redshift-dependent trends into our analysis.

\section{Results and analysis}
\label{sec:results}

\subsection{Variation in H$\alpha$ extinction with star formation
  rate, metallicity and stellar mass}
\label{sec:1dextinction}

\begin{figure}
  \centerline{\includegraphics[width=0.4\textwidth]{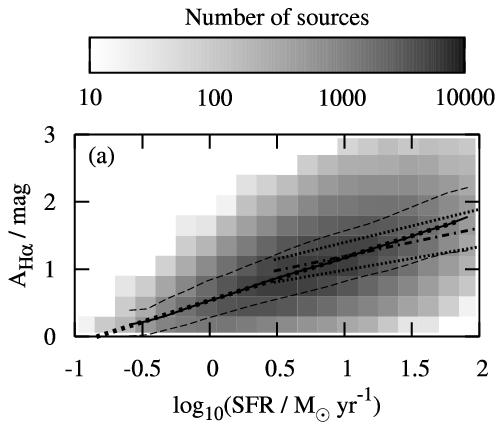}}
  \centerline{\includegraphics[width=0.4\textwidth]{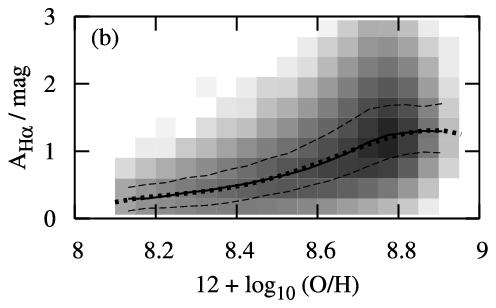}}
  \centerline{\includegraphics[width=0.4\textwidth]{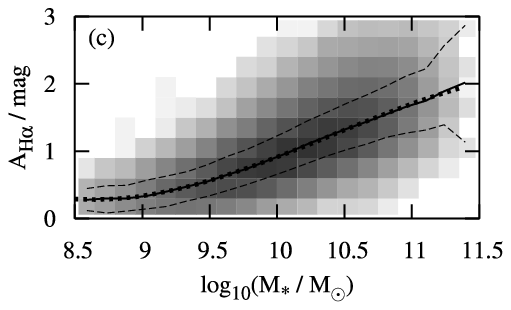}}
    \caption{The variation in H$\alpha$ extinction with (a) star-formation
      rate; (b) metallicity; (c) stellar mass.  The number of individual
      galaxies within each pixel is indicated by the grey-scale, and the
      binned data (median, $\pm$~$1\sigma$) by the solid and dashed lines
      respectively.  Polynomial fits to the median are shown by the dotted
      lines -- see Section~\ref{sec:model} for more details.  For
      comparison, the best-fitting relationship between SFR and extinction
      for $z=0.84$ galaxies from \citet{Garn09extinction} is shown on
      sub-figure (a), by the dash-dot and dotted lines (median and
      uncertainty respectively).}
    \label{fig:1dextinction}
\end{figure}

Fig.~\ref{fig:1dextinction}(a) shows the relationship between SFR and
extinction for the galaxies in our sample -- the grey-scale
illustrates the region of extinction / SFR space that is populated
with galaxies.  In order to quantify the observed trend, we separate
our sample into 20 logarithmically-spaced bins, selected by SFR.
Within each bin, we calculate the distribution of measured H$\alpha$
extinction values, which is overlaid (median, $\pm$~$1\sigma$).  A
clear increase is seen in the median extinction of galaxies with a
higher SFR, in agreement with previous studies of the low-redshift
Universe \citep[e.g.][]{Hopkins01,Sullivan01,Berta03}.

\citet{Garn09extinction} found that there was a well-defined relationship
between dust extinction and SFR for a sample of H$\alpha$-emitting
galaxies located at $z=0.84$.  Note that \citet{Garn09extinction}
calculated H$\alpha$ extinction using an independent method (a comparison
of H$\alpha$ and 24-$\mu$m SFR indicators, rather than from observations
of the Balmer decrement).  We overlay the \citet{Garn09extinction}
relationship on Fig.~\ref{fig:1dextinction}(a) -- the relationship is in
excellent agreement, with both the normalisation and the slope (despite
hints of a slight change in the latter) consistent to within $1\sigma$.
This confirms the conclusions of that work, that the dependence of dust
extinction on SFR does not change significantly between $z=0$ and $z=0.84$
(at least over the SFR range that was covered, 3 --
100~$M_{\odot}$~yr$^{-1}$).  This result also confirms that there should
be no significant difference in the typical extinction relations for
galaxies at different redshifts within our sample ($0.04 < z < 0.2$).

We repeat this analysis to determine the variation in typical
extinction with galaxy metallicity, which is shown in
Fig.~\ref{fig:1dextinction}(b). As has been found in previous studies
\citep[e.g.][]{Heckman98,Boissier04,Asari07}, the typical extinction
of a galaxy increases with metallicity and is approximately 1~mag for
solar-metallicity galaxies, which have $12 + {\rm log}_{10}({\rm O/H})
= 8.66$ \citep{Asplund04}.

Fig.~\ref{fig:1dextinction}(c) shows the equivalent results, binning
the sample by stellar mass.  The typical extinction of a galaxy
increases with stellar mass, as has been found in previous studies
\citep[e.g.][]{Brinchmann04,Stasinska04,Pannella09}, and rises
steadily between $10^{9}$ and $10^{11}$~$M_{\odot}$.  With the
exception of the outer bins of data (which contain fewer galaxies) the
typical 1$\sigma$ width of the distribution is $\sim0.3$~mag, comparable to the
distribution widths seen on the extinction / metallicity plot, and
substantially smaller than the $\sim0.5$~mag distribution width of the
extinction / SFR plot.

\subsection{Disentangling the relationships between H$\alpha$
  extinction and other galaxy properties} 
\label{sec:2d}

We have shown that the typical extinction of a galaxy correlates with its
SFR, metallicity and stellar mass.  In order to disentangle the complex
relationships between these properties, we separate our sample by one of
these properties into six sub-samples, and bin each sub-sample by both of
the remaining two variables.  For each bin, we calculate the distribution
of H$\alpha$ extinction (median $\pm1\sigma$).  The size of the SDSS DR7
dataset means that we can carry out this binning on a very fine scale --
in the following analysis, any bins containing fewer than 20 galaxies are
not shown for clarity, and contours illustrate the regions where bins
contain more than 100 or 1000 galaxies.

\begin{figure*}
  \centerline{
    \includegraphics[width=0.9\textwidth]{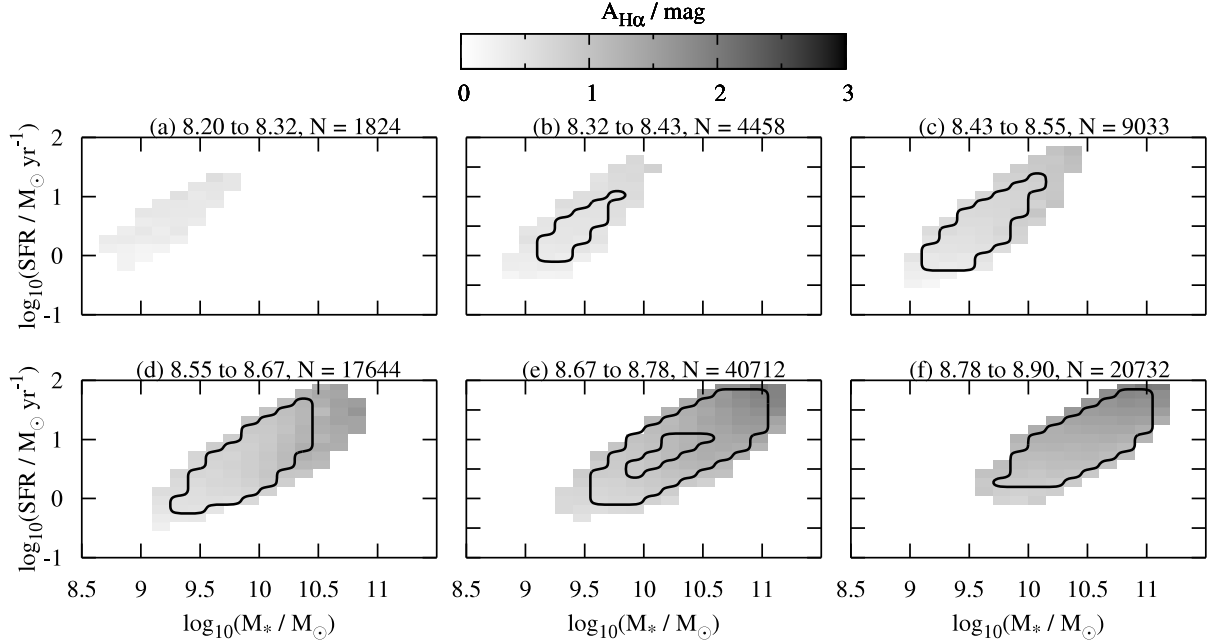}}
  \caption{The variation in H$\alpha$ extinction with stellar mass and
    star-formation rate, for sub-samples of galaxies having metallicity
    between $8.2 \leq 12 + {\rm log_{10}(O/H)} < 8.9$.  The grey-scale
    represents the median extinction within each bin, and contours
    illustrate the regions where bins contain more than 100 and 1000
    galaxies respectively.  The metallicity range and number of galaxies
    in each sub-sample are listed above each figure.}
  \label{fig:masssfrmetal}
\end{figure*}

We first take six sub-samples of galaxies with increasing metallicity,
and calculate how extinction varies with mass and SFR for each of
the sub-samples.  If changes in metallicity are the dominant factor
responsible for variations in the extinction of different galaxies,
then we would expect to see an approximately constant extinction in
each sub-sample, for any mass or SFR.  However,
Fig.~\ref{fig:masssfrmetal} shows the results: while different
metallicity galaxies populate different regions of the diagram, and
the typical extinction of a galaxy does increase for the
higher-metallicity sub-samples, there remains a significant variation
in extinction within each plot (in the sense that extinction increases
for higher masses or SFRs) that can not be explained purely through a
metallicity dependence.  The well-known correlation between stellar
mass and SFR of star-forming galaxies can be seen in
Fig.~\ref{fig:masssfrmetal} through the direction of the contoured
regions, and appears to be equally strong for galaxies of all
metallicities.

\begin{figure*}
  \centerline{
    \includegraphics[width=0.9\textwidth]{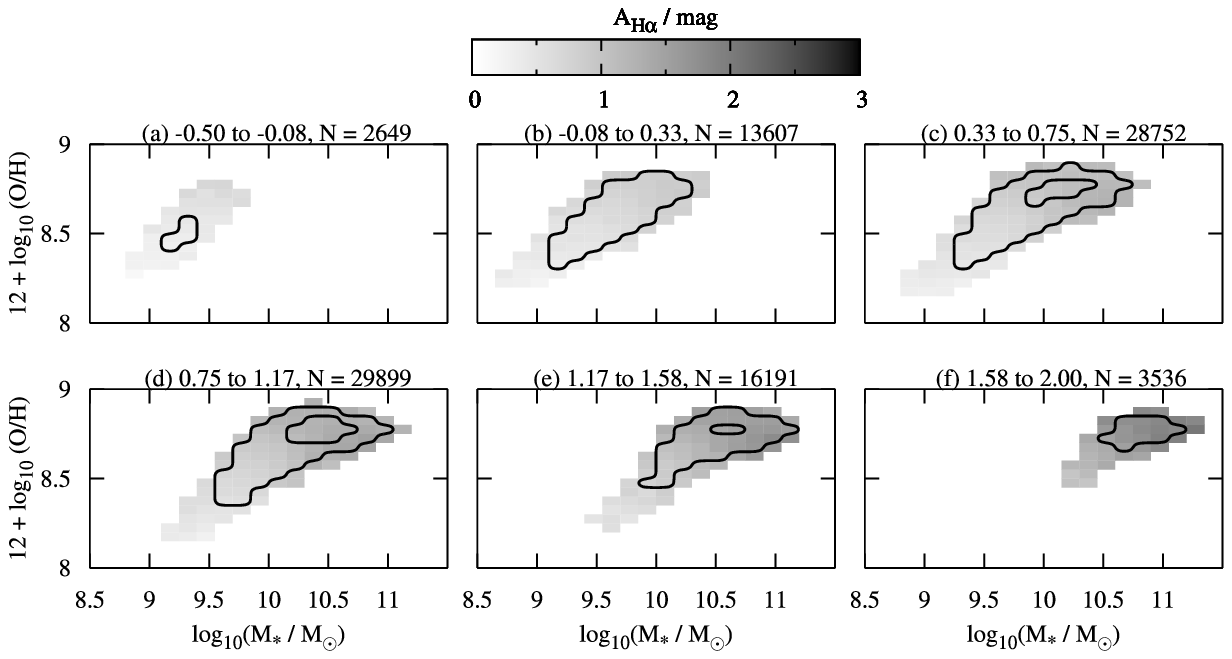}}
  \caption{The variation in H$\alpha$ extinction with stellar mass and
    metallicity, for sub-samples of galaxies having star-formation rate
    between $-0.5 \leq {\rm log}_{10}({\rm SFR}/M_{\odot}~{\rm yr}^{-1}) <
    2$.  The grey-scale and contours are as for
    Fig.~\ref{fig:masssfrmetal}, and the range of ${\rm log}_{10}({\rm
      SFR}/M_{\odot}~{\rm yr}^{-1})$ and number of galaxies in each
    sub-sample are listed above each figure. }
  \label{fig:massmetalsfr}
\end{figure*}

\begin{figure*}
  \centerline{
    \includegraphics[width=0.9\textwidth]{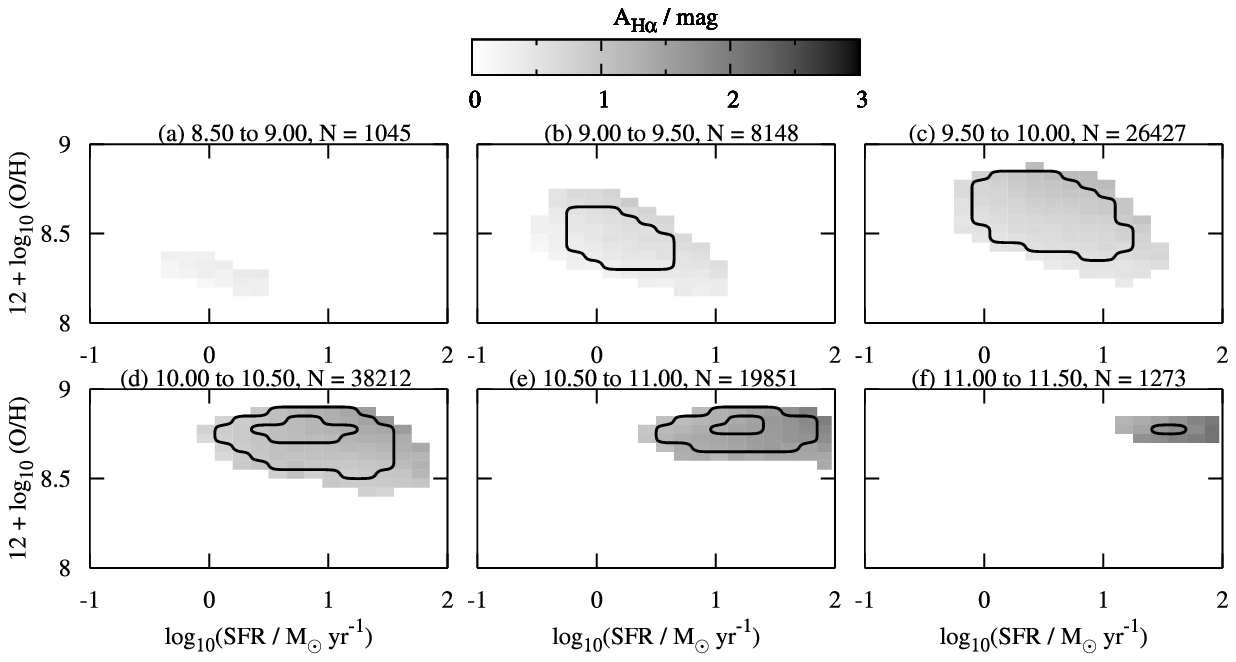}}
  \caption{The variation in H$\alpha$ extinction with star-formation rate
    and metallicity, for sub-samples of galaxies having stellar mass
    between $8.5 \leq {\rm log}_{10}(M_{*}/M_{\odot}) < 11.5$.  The
    grey-scale and contours are as for Fig.~\ref{fig:masssfrmetal}, and
    the range of ${\rm log}_{10}(M_{*}/M_{\odot})$ and number of galaxies
    in each sub-sample are listed above each figure. }
  \label{fig:sfrmetalmass}
\end{figure*}

Fig.~\ref{fig:massmetalsfr} shows the equivalent results, for
sub-samples of the data selected by SFR, and extinction as a function
of mass and metallicity.  The tight correlation between mass and
metallicity means that the width of the region that is populated in
each SFR sub-sample is relatively small.  \citet{Tremonti04} have
identified that the correlation begins to break down at stellar masses
above $10^{10.5}$~$M_{\odot}$, with more massive galaxies not
continuing to be more metal-rich -- this effect can be seen clearly in
Fig.~\ref{fig:massmetalsfr}.  As for Fig.~\ref{fig:masssfrmetal}, a
variation in the extinction of a galaxy within a given sub-sample can
be seen, implying that variations in SFR are not the dominant factor
responsible for differences in extinction between galaxies.

Fig.~\ref{fig:sfrmetalmass} shows the extinction as a function of SFR
and metallicity, for sub-samples selected by their stellar mass.  The
same effect as before is apparent -- different regions in the SFR /
metallicity plane are populated by galaxies in each of the stellar
mass sub-samples -- and a weak anti-correlation between SFR and
metallicity can be seen, for galaxies of a particular stellar mass.
However, in contrast to the previous two figures, there is little
variation in the median extinction of a galaxy at any populated point
in each of the sub-samples.  This implies that the majority of the
variation in extinction between galaxies is driven by variations in
stellar mass, rather than SFR or metallicity.

\subsection{Principal component analysis}

In order to quantify the relative importance of each of the parameters
that we are considering, we normalise each to have zero mean and unit
standard deviation, and calculate the covariance matrix of the
resulting variables.  Each entry in this matrix measures the strength
of the linear dependence between two variables, and ranges between
$+1$ (a perfect correlation between the variables), 0 (no correlation)
and $-1$ (perfect anti-correlation).  Table~\ref{tab:covariance}
reports the results: mass is strongly correlated with all the other
variables, and SFR and metallicity are also strongly correlated
with extinction.

\begin{table}
  \begin{center}
    \caption{The covariance matrix for the star-forming galaxies in
    our sample.  All parameters have been normalised to have a
    distribution with zero mean and unit standard deviation.  `SFR'
    $\equiv$ log$_{10}$(SFR / $M_{\odot}$~yr$^{-1}$), `Mass' $\equiv$
    log$_{10}(M_{*} / M_{\odot})$, `Metallicity' $\equiv$ $12 + {\rm
    log}_{10}({\rm O/H})$, `Extinction' $\equiv$ $A_{\rm H\alpha}$.}
    \label{tab:covariance}
    \begin{tabular}{c|cccc}
            & `SFR' & `Mass' & `Metallicity' & `Extinction'\\
\hline
`SFR'         & 1.000 & 0.734 & 0.316 & 0.612\\
`Mass'        & 0.734 & 1.000 & 0.717 & 0.719\\
`Metallicity' & 0.316 & 0.717 & 1.000 & 0.572\\
`Extinction'  & 0.612 & 0.719 & 0.572 & 1.000\\
\end{tabular}
\end{center}
\end{table}

We use principal component analysis \citep[PCA; e.g.][]{Boroson92} to
quantify the relationships between these variables.  This technique
creates eigenvectors that are made up of linear combinations of the
input variables, and are rotated to span the directions of maximum
variance in the data.  Principal component (PC) 1 contributes
71~per~cent of the total variance, with weightings for the normalised
SFR, mass, metallicity and extinction parameters as listed in
Table~\ref{tab:pca} -- i.e.\ the majority of the total variance in our
sample can be explained simply by all four variables being correlated
with each other, as has been found in previous sections.  If all of
the normalised quantities contributed equally to this PC, then they
would all have weights of $1/\sqrt{4} = 0.5$; we find that mass
contributes slightly more than metallicity or SFR to the first PC, but
that all are broadly equivalent.

\begin{table}
  \begin{center}
    \caption{The four principal components, along with the percentage
    of variance in the data that they explain.  Parameters are as
    described in Table~\ref{tab:covariance}.}
    \label{tab:pca}
    \begin{tabular}{ccccc}
 & `SFR' & `Mass' & `Metallicity' & `Extinction'\\
\hline
PC1 (71\%) & 0.469    & 0.559    & 0.453    & 0.512\\
PC2 (17\%) & $-0.679$ & 0.011    & 0.733    & $-0.040$\\
PC3 (9\%)  & 0.313    & 0.331    & 0.237    & $-0.858$\\
PC4 (3\%)  & $-0.471$ & 0.760    & $-0.448$ & $-0.002$\\
\end{tabular}
\end{center}
\end{table}

PC2 contributes a further 17~per~cent of the total variance, with
significant weightings of $-0.679$ for SFR and 0.733 for metallicity
(see Table~\ref{tab:pca}) -- i.e.\ the majority of the remaining
scatter in our data can be attributed to an anti-correlation between
SFR and metallicity, for little variation in mass and extinction.  A
weak anti-correlation can be seen in each of the panels of
Fig.~\ref{fig:sfrmetalmass} (as illustrated by the contoured regions).
However, we have not fully removed the effects of stellar mass in
these panels, as there is still a 0.5~dex variation in stellar mass
across each of the sub-samples.  Note that since the weightings of
mass and extinction in PC2 are very small, the anti-correlation
between SFR and metallicity will partially cancel out the correlation
between SFR and metallicity from PC1.

The direction of PC2 implies that, if the stellar mass dependencies of
other variables are removed from our dataset, we should see an
anti-correlation between SFR and metallicity.  In order to remove
these dependencies, we bin the full dataset by stellar mass and fit a
polynomial to the median mass and SFR for each bin, which gives
\begin{equation}
{\rm log}_{10}({\rm SFR}/M_{\odot}~{\rm yr}^{-1}) = 0.64 +0.76~X +
0.12~X^{2},
\label{eq:masssfr}
\end{equation}
where $X = {\rm log}_{10}(M_{*}/10^{10} M_{\odot})$.  

We repeat this with the binned median mass and metallicity, to obtain
\begin{equation}
12 + {\rm log}_{10}({\rm O/H}) = 8.68 + 0.23~X - 0.13~X^{2},
\label{eq:massmetal}
\end{equation}
In both cases, moving to a higher-order polynomial did not
significantly improve the fit.

\begin{figure}
  \begin{center}
    \includegraphics[width=0.4\textwidth]{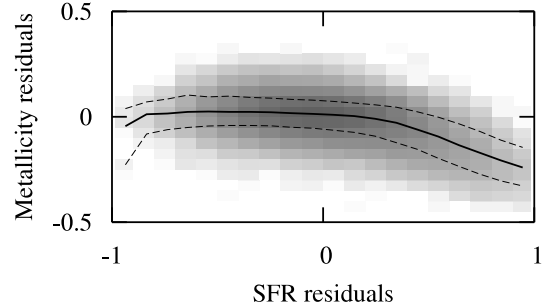}
    \caption{The distribution of SFR and metallicity residuals, after
    stellar mass dependencies are removed using
    equations~(\ref{eq:masssfr}) and (\ref{eq:massmetal}).  SFR
    residuals are given by log$_{10}$(SFR/SFR$_{\rm model}$), and
    metallicity residuals by [$12 + {\rm log}_{10}({\rm O/H})$] --
    [$12 + {\rm log}_{10}({\rm O/H})_{\rm model}$].  The distribution
    of residuals is indicated by the grey-scale, and the solid and
    dashed lines indicate how the metallicity residual varies with SFR
    residual (median, $\pm$~$1\sigma$).}
    \label{fig:massresiduals}
  \end{center}
\end{figure}

We use these relations to predict the SFR and metallicity of all galaxies
in the sample (the redshift range studied is sufficiently small that any
cosmic evolution of these relations is negligible).  By subtracting these
predicted values from the observed values, we obtain the amount of SFR or
metallicity variation that is not related to the overall stellar mass
dependency, which we denote as the SFR and metallicity residuals.  The
relationship between SFR and metallicity residuals is shown in
Fig.~\ref{fig:massresiduals}.  An anti-correlation between SFR and
metallicity, after stellar mass effects are removed, appears in the data
for positive SFR residuals.  Note that PCA only finds linear relationships
between variables -- as we have had to fit quadratics in order to
represent the data successfully, we do not recover a pure anti-correlation
between the two residuals.

\citet{Ellison08} have found a similar trend that for a given stellar
mass, galaxies with a high SFR have systematically lower metallicities
than their low-SFR counterparts.  They also find that more
concentrated galaxies, with a smaller half-light radius, tend to be
more metal-rich.  Their interpretation of these results is that
galaxies with a high surface mass density may have had a higher star
formation efficiency in the past than their lower-density counterparts
-- this would lead to an increased production of metals, but lower
current SFRs as more of the gas reservoir would have been depleted.
This interpretation is consistent with our results.

PC3 contributes 9~per~cent of the total variance, with the only
significant weighting being for extinction.  This may imply that there is
a residual scatter in the extinction of each galaxy that is caused by some
property other than mass, SFR or metallicity.  PC4 contributes the
remaining 3~per~cent of the variance, and will be dominated by noise in
the sample.

\subsection{Modelling the variation in extinction}
\label{sec:model}
Having shown that extinction is correlated with various properties of
a galaxy, we now attempt to model these variations, and identify the
amount of information that is required in order to accurately estimate
the typical extinction of a galaxy.  We will quantify the accuracy of
our models in two ways:
\begin{enumerate}
\item calculating the residuals between the model and observed
  extinctions for all galaxies in our sample, and measuring the width
  of the Gaussian distribution that describes these residuals;
\item plotting the distribution of residuals against galaxy mass, SFR
  and metallicity, in order to identify any significant trends that
  have not been accounted for in the models.
\end{enumerate}

\begin{figure*}
  \begin{center}
    \includegraphics[width=0.8\textwidth]{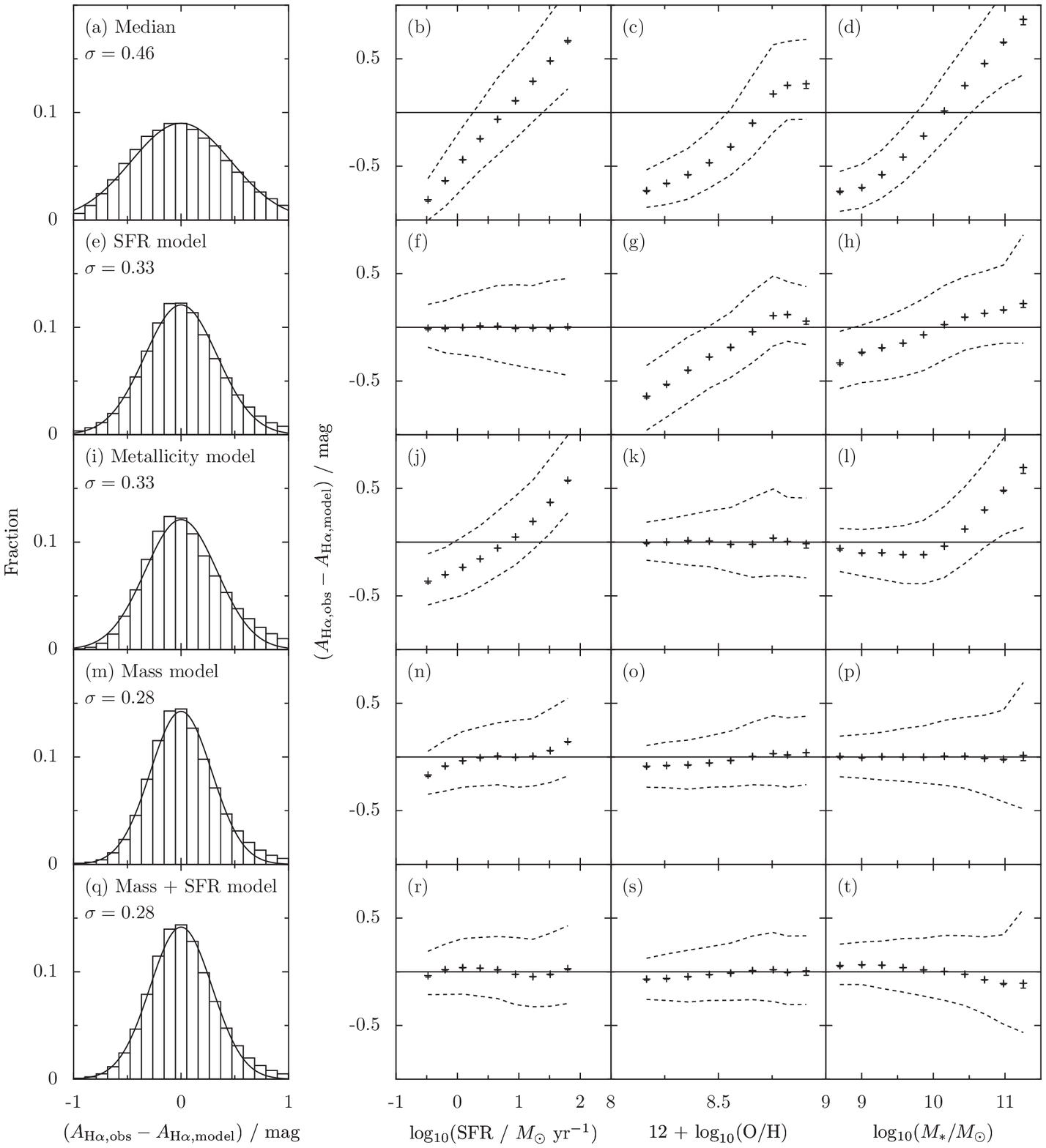}
    \caption{The distribution of residual extinctions for our sample,
    for each of the models described in Section~\ref{sec:model}.  Each
    row represents a different model, which is named in the left-hand
    panel.  The first column shows the overall distribution of
    residuals, along with the best-fitting Gaussian centered on zero.
    Column~2 shows the variation in residuals with SFR, column~3 shows
    the variation with metallicity and column~4 shows the variation
    with stellar mass.  Points with error bars indicate the median and
    error in the median; dashed lines indicate the $\pm1\sigma$
    distribution width.}
    \label{fig:residuals}
  \end{center}
\end{figure*}

Our first model is the assumption that all galaxies can be represented by
a single value of extinction, which we take to be the median value of
$A_{\rm H\alpha}$ in our sample, 1.03~mag.  Fig.~\ref{fig:residuals}(a)
shows that the distribution of residuals is roughly Gaussian, with width
$\sigma=0.46$~mag.  The strong variations in extinction with SFR,
metallicity or stellar mass are shown in Figs.~\ref{fig:residuals}(b) to
(d), as previously identified in Section~\ref{sec:1dextinction}.  We show
both the uncertainty in the median estimate, and the $\pm1\sigma$
distribution widths, in order to fully quantify the distribution.  This
model is the one most commonly used in the literature for
H$\alpha$-selected samples, but is clearly not a good fit to the data.

Our next three models come from polynomial fits to the binned data
from Section~\ref{sec:1dextinction}, in the form of
\begin{equation}
A_{\rm H\alpha} = \sum_{i = 0}^{n} B_{i}~X^{i}.
\label{eq:extinctionmodel}
\end{equation}
The extinction / SFR relationship, with $X ={\rm log}_{10}({\rm
  SFR}/M_{\odot}~{\rm yr}^{-1})$, is well-fitted by a linear relationship.
The extinction / metallicity relationship, with $X = [12 + {\rm
    log}_{10}({\rm O/H})] - 8.5$, requires a much higher-order polynomial
(5th order) in order to model the curvature at high metallicities.  We
note here that it is unlikely that an extinction / metallicity
relationship would be of much practical use -- if sufficient spectroscopy
exists that a reliable metallicity can be calculated, it would be better
to measure extinction values directly.  However, we include this model in
our discussion for completeness.  Finally, the extinction / stellar mass
relationship, with $X = {\rm log}_{10}(M_{*} / 10^{10}M_{\odot})$, can be
modelled by a 4th-order polynomial.  These relationships are overlaid on
Fig.~\ref{fig:1dextinction} for reference, and all are a good fit to the
binned data -- we do not obtain a significantly better fit by moving to
higher-order polynomials. The co-efficients of the polynomial fits are
given in Table~\ref{coeffs}.

\begin{table}
\caption{\label{coeffs}Co-efficients of the polynomial fits to the
  extinction as determined using Eq.~\ref{eq:extinctionmodel}.}
  \begin{center}
\begin{tabular}{lccccc}
Model & $B_{0}$ & $B_{1}$ & $B_{2}$ & $B_{3}$ & $B_{4}$ \\ 
\hline
SFR         &  0.53 & 0.54 &      &       &       \\
Metallicity &  0.64 & 1.75 & 2.96 & -2.57 & -12.4 \\
Mass        &  0.91 & 0.77 & 0.11 & -0.09 &       \\
$\mu$       &  1.11 & 0.69 & -0.19& -0.18 &       \\
\end{tabular}
\end{center}
\end{table}

The overall distributions of residuals are shown in
Fig.~\ref{fig:residuals}(e), (i) and (m), and again are consistent with
being Gaussian, with zero mean and widths of 0.33, 0.33 and 0.28~mag for
SFR, metallicity and mass.  All three models show significant improvement
over the assumption of a single extinction for all galaxies.  The
contribution to these residuals from uncertainties in measuring the
H$\alpha$ and H$\beta$ fluxes is about 0.2~mag, from
Fig.~\ref{fig:properties}(f), which is about half of the total if errors
are independent, and added in quadrature. Measurement errors in the SFR,
metallicity and stellar mass parameters will account for some of the
remaining scatter, but are unlikely to account for all. This suggests that
there remains up to 0.2~mag of intrinsic scatter between different
galaxies (cf. PC3 in Section~3.3).

The distribution of residuals for each of these models is shown in
Fig.~\ref{fig:residuals} as a function of SFR, metallicity and stellar
mass, in order to test for remaining trends.  By construction, the SFR
residuals show no trend when the SFR model is used, and equivalent results
are found for the other two models.  However, the SFR model leads to a
considerable variation in the residual extinction with either metallicity
or stellar mass, in the sense that the model over-predicts the extinction
of low-metallicity or low-mass galaxies.  A similar result is found for
the metallicity model, with extinction being over-predicted for low-SFR or
low-mass galaxies.  However, the mass model shows significantly smaller
residual trends, as expected from the results of Section.~3.2 and
Figure~7. This implies that it describes the data with less bias than the
alternatives, although some weak trends with SFR and metallicity do still
remain.

We attempt to improve upon the mass model, by adding a further
SFR-dependent term to equation~(\ref{eq:extinctionmodel}) with
coefficients of $B_{0} = -0.09$ and $B_{1} = 0.13$ (the best-fit values
after taking out the mass dependence); this will produce an improvement if
the scatter around the SFR-mass relation correlates strongly with
extinction.  The results are shown in the bottom row of
Fig.~\ref{fig:residuals} -- a slight improvement in the variation of
residuals with SFR is seen (by construction), but the residuals show a
worse trend with stellar mass than the previous model (since the strong
SFR-mass correlation means that the additional SFR fit worsen the mass
fit); the overall Gaussian width remains unchanged, at 0.28~mag.  The
largest residuals are seen at high stellar masses, possibly because this
is the region where the mass / SFR correlation begins to break down, due
to the effects of downsizing (where the most massive galaxies have
already formed their stars, and undergo smaller amounts of current star
formation than their less-massive counterparts).  Concurrent fitting of
the SFR and mass polynomials (even with the inclusion of cross-terms) was
also attempted but does not decrease the overall Gaussian width of the
residuals: although the trends of the residuals with SFR and mass improve
slightly, those with metallicity worsen. Equivalent results are found if
either a metal-dependent term is added to
equation~(\ref{eq:extinctionmodel}), or both a SFR-dependent and
metal-dependent term are added, or all three terms are fitted concurrently
-- in all cases, the residuals are consistent with a Gaussian having a
width of 0.28~mag. These results confirm that knowledge of stellar mass is
sufficient to model the extinction of a galaxy in a statistical sense --
once the mass variation is removed, very little improvement is obtained by
a more complex model.

Finally, we note that very recently \citet{Mannucci10} have proposed a
`fundamental metallicity relation', whereby galaxies define a tight
surface in the 3-dimensional space of mass, SFR and metallicity. They
introduce a new quantity, $\mu = \rm{log}_{10}$(Mass) - 0.32
$\rm{log}_{10}$(SFR), against which they argue that metallicity scatter of
nearby galaxies is minimised. To test whether this can improve extinction
estimates, we model the median extinction as a function of $\mu$ using a
4th-order polynomial (with co-efficients given in Table~\ref{coeffs}).
However, we find that the resultant Gaussian width of the residuals
relative to this model is 0.31~mag, with residual trends remaining with
each of mass, SFR and metallicity that are worse than those of the mass
model. We conclude that, for modelling extinction, the stellar mass alone
provides a better (and simpler) input than $\mu$.

\subsection{H$\alpha$ S/N bias}

One of our primary selection requirements was that sources have an
H$\alpha$ S/N ratio greater than 20, as measured from the observed line
flux. By doing this, we are implicitly biasing our sample against highly
extinguished sources -- the error on the line flux measurement will be
independent of the extinction of a galaxy, while the observed flux will be
lower for galaxies with higher extinction.

In order to confirm that this bias does not significantly influence our
conclusions, we repeat our analysis using an alternative selection
procedure which is less prone to an extinction bias. For galaxies in which
the H$\alpha$ and H$\beta$ emission lines are both detected with S/N ratio
greater than 3, we use equation~(\ref{eq:extinction}) to calculate the
H$\alpha$ extinction.  Similarly, for those galaxies with H$\alpha$ S/N
ratio above 3, but H$\beta$ S/N ratio below 3, we estimate the extinction
using the 3$\sigma$ limit on the H$\beta$ flux (strictly this is a lower
limit on the extinction). We then correct the H$\alpha$ line fluxes for
the calculated extinction, and derive the S/N ratio with which the
H$\alpha$ line would have been detected in the absence of extinction. We
then set a primary selection requirement that this unextinguished
H$\alpha$ S/N ratio should be greater than 75, and we repeat the analysis
of the paper. This selection procedure does not completely remove the
effects of extinction bias (the most extinguished sources may still be
excluded if their observed H$\alpha$ S/N ratio is below 3, or if their
H$\beta$ line is undetected leading to an underestimated extinction and
undercorrection of their H$\alpha$ flux), but it does greatly reduce the
bias. Therefore, if the same trends are found, that would provide
confidence in their reliability.

The median extinction of H$\alpha$ in the new sample is 1.35 magnitudes,
significantly higher than the previous selection. This increase is driven
both by the exclusion of low-extinction galaxies (mostly at lower masses;
these were in the original sample but fail to reach the higher threshold
for the revised sample) and by the inclusion of additional high-extinction
galaxies (especially at higher masses). We re-calculate the relations with
extinction determined in Section~\ref{sec:model}. The fitted polynomial
parameters for the mass / extinction relationship are $B_{0} = 1.09$,
$B_{1} = 0.65$, $B_{2} = 0.12$ and $B_{3} = -0.031$; this has almost
exactly the same shape as the previous relation, but is shifted up by
about 0.18 magnitudes of extinction. Equivalent shifts are seen in the
extinction relationships with other parameters. The distributions of
extinction residuals (cf.  Fig.~\ref{fig:residuals}) again show similar
trends, with a scatter of 0.40 around the median extinction value being
reduced to 0.37, 0.34 and 0.28 when taking out the SFR, metallicty and
mass models, respectively.  This confirms that mass is the predominant
factor influencing the extinction of galaxies; once again, adding in
additional SFR- or metallicity-dependent terms does not further decrease
the scatter.

These tests leads us to conclude that although we are excluding some of
the more highly extinguished sources from our sample, the trends
determined from the analysis are robust.

\section{Discussion and Conclusions}
\label{sec:discussion}
We have used the superb spectroscopic coverage of SDSS DR7 to
disentangle the effects of three of the physical parameters that are
known to correlate with the dust extinction of a galaxy.  We select a
clean sample of star-forming galaxies using the BPT diagram, and
calculate extinctions from observed Balmer decrements.  We calculate
metallicities from the O3N2 indicator, and after correcting for dust
extinction and aperture size, we calculate SFRs from the H$\alpha$
line.  Stellar mass estimates have been calculated independently by
the MPA/JHU group.

We confirm that the typical dust extinction of a galaxy increases with
greater SFR, metallicity or stellar mass.  By selecting various
sub-samples of galaxies, we show that the dominant factor is stellar
mass -- we conclude that the observed correlations between extinction
and SFR, or extinction and metallicity, are largely secondary effects
brought about by the known dependence of these parameters on the
stellar mass of a galaxy.  The implications that we draw from our
results are:

\begin{itemize}
\item As galaxies build up their stellar mass over time, they also
  build up their dust content.  This increase in dust leads to a
  greater H$\alpha$ extinction, and therefore a dependence of
  extinction on stellar mass is seen.

\item Massive galaxies are able to retain a greater fraction of the
  metals that they produce, due to their deep potential wells, while
  less-massive galaxies lose metals from their shallower potential
  wells through the effects of galactic winds
  \citep[e.g.][]{Tremonti04}.  A relationship between metallicity and
  extinction is therefore formed -- this is a secondary effect brought
  about by the dependence of both extinction and metallicity on
  stellar mass.

\item More massive galaxies are capable of forming a greater amount of
  stars in a starburst, and therefore a relationship between the SFR
  and extinction of a galaxy is set up -- this is also a secondary
  effect brought about by the relationship between SFR and stellar
  mass.  This relationship will only hold for galaxies that are
  currently forming stars, and will break down at the highest masses,
  since these galaxies formed their stars at earlier epochs
  \citep{Cowie96}.  However, extremely high-mass galaxies are poorly
  represented in our sample, since it was selected by the presence of
  H$\alpha$ emission to be undergoing current star formation.

\item An anti-correlation is seen between SFR and metallicity, for
  galaxies of a given stellar mass.  This may be due to variations in
  star formation efficiency between galaxies, as discussed by
  \citet{Ellison08}, where galaxies that have formed more stars in the
  past have a lower current SFR, and higher metallicity.
\end{itemize}

Using equation~(\ref{eq:extinctionmodel}) it is possible to predict the
extinction of a galaxy with a given stellar mass, to within a typical
error of 0.28~mag (note that as discussed in Section~3.5, samples selected
with different selection effects may require small offsets in
normalisation of this equation, but the functional form and the scatter
around it is largely robust to changes in selection technique). This
typical error is broadly comparable to the accuracy with which extinctions
can be estimated from the Balmer decrement, for our sample of galaxies
selected with an H$\alpha$ S/N ratio $>20$, although this comparison will
of course depend upon the S/N ratio of available H$\alpha$ and H$\beta$
observations.  We believe that the stellar mass of a galaxy is the best
indicator to use when estimating the typical amount of dust extinction,
both because it appears to be the fundamental parameter that governs dust
extinction, and also because it can be estimated in a way that is almost
unaffected by dust, using the rest-frame $K$-band luminosity alone
\citep[e.g.][]{Longhetti09}.  For some applications, it might be simpler
to use the SFR-dependent correction (for example, correcting an observed
H$\alpha$ or UV luminosity function for the effects of dust).  However, we
caution that this form of correction has the potential for introducing a
significant mass-dependent bias into the results, and should be used with
care.

Estimating the extinction of truly `passive' galaxies (with SFR
$<0.1$~$M_{\odot}$~yr$^{-1}$) is outside the scope of this work, as these
are not detected in our H$\alpha$ selected samples. However, for massive
galaxies, we note that the H$\alpha$ selection limit corresponds to star
formation rates well over an order of magnitude below their historical
average. These galaxies display high levels of extinction, implying that
high-mass `passive' galaxies are also likely to have high levels of
extinction of the small amount of H$\alpha$ emission that they do produce.

We note that although all of our results have been derived for
H$\alpha$ extinction, they can be generally applied to line emission
at another wavelength through the use of
equation~(\ref{eq:extinction}) and the \citet{Calzetti00} dust
attenuation law.  If the extinction of UV radiation is required,
rather than line emission, then a factor of 0.5 needs to be applied to
this equation -- see \citet{Calzetti00} and \citet{Garn09extinction}
for further details.

Estimates of star-formation at high redshift ($z>1$) are commonly made
using a wide range of different estimators: UV continuum; [OII] or
H$\alpha$ emission lines; sub-mm, mid-IR or far-IR continuum; radio
emission. Each of these has their own advantages and disadvantages, with
different sensitivities, angular resolutions, reliability of calibration,
and effects of dust extinction \citep[cf.][]{Kennicutt98}. The latter
effect is particularly important for star-formation rates estimated from
H$\alpha$ or UV selected samples (the most sensitive selection methods
currently available); previous star-formation rate estimates from such
samples have either required assumptions about the global extinction of
all galaxies (e.g. a constant -- a hypothesis which can be ruled out by
the results in this work), spectroscopic observations of each galaxy to
measure individual dust extinctions (time-consuming, and technically
challenging at high redshift), or a sufficiently large number of
photometric data-points to enable SED-fitting techniques to estimate
extinctions without issues of degenerate solutions. The results presented
in this work suggest that a representative extinction correction can be
derived from the mass, which can be estimated using far fewer photometric
data points. Provided that the relation between mass and extinction is
confirmed to be independent of redshift, this result could have important
applications for studying large samples of star-forming galaxies,
especially at high redshifts.

\section*{Acknowledgements}
This paper is dedicated to the memory of, and in tribute to, Timothy
Garn. Timothy had largely completed the analysis and writing of the paper
before his untimely death, and PNB put the finishing touches to it. PNB is
grateful for support from the Leverhulme Trust.  The authors thank Jarle
Brinchmann for providing advice on using the MPA/JHU catalogues, and
making the duplicate source list available. PNB thanks the anonymous
referee for helpful comments.

\setlength{\labelwidth}{0pt}
\bibliography{/home/tsg/data/Documents/Papers/References}
\bibliographystyle{/home/tsg/data/Documents/Papers/mn2e}
\label{lastpage}

\end{document}